\documentclass[11pt]{article} 
\usepackage[dvips]{graphics}
\usepackage[dvips]{color}
\usepackage{epsfig}
\usepackage{psfig}
\usepackage{lscape}
\usepackage{graphicx}
\newcommand{\beq}{\begin{equation}}
\newcommand{\eeq}{\end{equation}}
\newcommand{\beqn}{\begin{eqnarray}}
\newcommand{\eeqn}{\end{eqnarray}}
\newcommand{\bi}{\begin{itemize}}
\newcommand{\ei}{\end{itemize}}

\newcommand{\RR}{{\rm I\kern -.2em  R}}

\newcommand{\half}{{1\over2}}

\def\lsi{\raise0.3ex\hbox{$<$\kern-0.75em\raise-1.1ex\hbox{$\sim$}}}
\def\gsi{\raise0.3ex\hbox{$>$\kern-0.75em\raise-1.1ex\hbox{$\sim$}}}

\newcommand{\J}[4]{{#1} {\bf #2} (#3) #4}

\newcommand{\NP}{Nucl.~Phys.}
\newcommand{\NPSup}{Nucl.~Phys.~B (Proc.~Suppl.)}
\newcommand{\PL}{Phys.~Lett.}
\newcommand{\PR}{Phys.~Rev.}

\newcommand{\CPC}{Comput.~Phys.~Commun.} 
\newcommand{\JCP}{J.~Compt.~Phys.} 
\makeatletter \@addtoreset{equation}{section} \makeatother

\makeatletter
\renewcommand\section{\@startsection {section}{1}{\z@}%
                                   {-5.5ex \@plus -1ex \@minus -.2ex}% bfr-skip
                                   {2.3ex \@plus.2ex}%
                                   {\normalfont\large\bfseries}}
\renewcommand\subsection{\@startsection{subsection}{2}{\z@}%
                                     {-3.25ex\@plus -1ex \@minus -.2ex}%
                                     {1.5ex \@plus .2ex}%
                                     {\normalfont\normalsize\bfseries}}
\renewcommand\thesection {\@arabic\c@section}
\renewcommand\thesubsection   {\thesection.\@arabic\c@subsection}
\renewcommand{\@seccntformat}[1]{%
\csname the#1\endcsname.\hspace{1.0em}}
\makeatother

\begin{document}

\begin{titlepage}
\begin{flushright}
CERN-TH/2003-62 \\
DESY 03-029 \\
%hep-lat/\\
\end{flushright}
\begin{centering}
\vfill
 
{\Large{\bf Reducing Residual-Mass Effects for Domain-Wall Fermions}}

\vspace{0.8cm}

Karl Jansen$^{\rm a}$ and 
Kei-ichi Nagai$^{\rm b}$

\vspace{0.8cm}

{\em $^{\rm a}$%
NIC/DESY Zeuthen, Platanenallee 6, D-15738 Zeuthen, Germany\\}

\vspace{0.3cm}

{\em $^{\rm b}$%
CERN, Theory Division, CH-1211 Geneva 23, Switzerland\\}

\vspace*{0.8cm}

\end{centering}
 
\vspace*{0.4cm}

\noindent
\begin{abstract}
It has been suggested to project out 
a number of low-lying eigenvalues 
of the four-dimensional Wilson--Dirac operator
that generates the transfer matrix 
of domain-wall fermions
in order to improve simulations with domain-wall fermions.
We investigate 
how this projection method reduces the residual chiral symmetry-breaking
effects for a finite extent of the extra dimension.
We use the standard Wilson as well as 
the renormalization--group--improved gauge action. 
In both cases we find a substantially reduced residual mass 
when the projection method is employed. 
In addition, the large fluctuations in this quantity disappear.  
\end{abstract}
\vfill
\noindent

%\noindent
%PACS numbers: 
%
%11.15.Ha, %        Lattice gauge theory
%11.30.Rd, %        Chiral symmetries
%\\
%Keywords:

\vspace*{1cm}
 
%\noindent
%CERN-TH/2003-\\

\vfill

\end{titlepage}

%%%%%%%%%%%%%%%%%%%%%%%%%%%%% SECTION %%%%%%%%%%%%%%%%%%%%%%%%%%%%%%%%%%%
\section{Introduction}
\label{sec:intro}

Domain-wall fermions (DWF) preserve 
chiral symmetry \cite{kaplan,shamir,fursha}
when the lattice size in the 5th direction, $N_s$, 
is taken to infinity.
The approach to the chiral limit
is exponential in $N_s$,
with a rate given by the eigenvalues of the transfer matrix 
along the 5th direction,
which is a local operator 
in 4 dimensions \cite{overlap,kikukawanoguch,kikukawa,borici}.
A measure of chiral symmetry breaking, taking place for finite $N_s$,  
is the residual mass, $m_\mathrm{res}$, derived from 
the axial Ward--Takahashi identity. 

Even if the restoration of chiral symmetry 
is expected to be exponentially fast in $N_s$,
in practice $m_\mathrm{res}$ can decrease very slowly 
as first shown by the CP--PACS collaboration \cite{cppacs,eigenv}.
The slow convergence of the residual mass is due to
the existence of very small eigenvalues 
of the four-dimensional operator defining 
the transfer matrix along the 5th direction.
In particular, 
at large $N_s$ these low-lying modes dominate the convergence rate
\cite{eigenv} and render 
the recovery of chiral symmetry difficult.
Even if the residual mass is very small,
it is then not clear 
whether and what distortions of chiral symmetry are still present.
Since large numerical simulations with DWF are being performed  
(see e.g. refs. \cite{cppacs,rbc} and the reviews \cite{vranas,pilar})
it becomes important to find 
ways around this obstacle.    
Such solutions for  
improving the chiral properties of DWF
then have to come from eliminating these low-lying modes.

One idea to reduce these small eigenvalues is the 
improvement of the gauge actions \cite{cppacs,eigenv,rbc,dbw2} 
such as Iwasaki \cite{iwasakiaction} or DBW2 \cite{dbw2action}.
However, 
besides the potential difficulties 
with unitarity violations \cite{necco} 
and the sampling problems of topological charge sectors \cite{newrbc},
this method does not solve the problem completely.
For example,  
with the Iwasaki gauge action, 
the convergence rate also becomes slow at large $N_s$ \cite{eigenv}.
The reason is that again 
very small eigenvalues of the transfer matrix appear in this case,
though less frequently than for the Wilson gauge action.
Using the DBW2 gauge action seems to be much better in this respect 
\cite{rbc,newrbc},
but it is unclear whether these small eigenvalues could eventually
appear there, too, 
leading to similar problems.
A perturbative analysis \cite{shamirpert} suggests a modification of 
the four-dimensional component of the domain wall operator to tackle
the problem. This is, however, not yet tested in simulations. 

Another method to eliminate the disturbing effect of the small eigenvalues 
and the corresponding set of eigenstates of the transfer matrix
is to project them out and lift them in a way that does not change 
the $N_s \rightarrow \infty$ limit of the DWF operators \cite{project,boudp}.
In this paper,
we investigate the projection method based on ref. \cite{project},
where the projection is performed in the transfer matrix itself.
In ref. \cite{boudp}, an alternative projection is implemented through a modification of the boundary terms. 
The philosophy of both approaches is the same as 
the one using the transfer matrix.
The aim of this article is to investigate the effects 
of the projection method on the residual mass 
in quenched simulations. As we will see, the projection method 
works very well, 
leading to a substantial improvement in the residual mass. 

Let us emphasize that
simulations with DWF can be considered
under two aspects.
The ``purist's'' approach demands exact chiral symmetry
at non-zero lattice spacing.
Here {\em any} violation of chiral symmetry
(in practice up to machine precision)
is not tolerable.
Hence the value of $N_s$ is to be taken 
as large as possible and the additive mass renormalization
$m_\mathrm{res}\ne 0$ has to be eliminated.
Thus the projection method discussed here,
or any method leading to the same improvement, 
becomes an unavoidable necessity
in this case.
A different, more 
practical point of view
is to consider DWF 
at finite, and even small,
values of $N_s$ as a highly improved Wilson fermion.
Also in this case,
the projection method will accelerate the numerical simulation considerably
and should therefore be employed.
%

%%%%%%%%%%%%%%%%%%%%%%%%%%%%%%% SECTION %%%%%%%%%%%%%%%%%%%%%%%%%%%%%%%%%%%%%
%\clearpage
\section{Domain-wall fermions and Ward--Takahashi identity}
\label{sec:dwf}

In this section,
we establish our notation and give the Ward--Takahashi identity 
in order to define the residual mass.
For completeness, we give here the definition of 
the domain-wall operator and its relation to the 4D operator 
satisfying the Ginsparg--Wilson
equation \cite{gw}.
We follow the presentation of \cite{project}.
Derivations of this formulae can be found 
in \cite{shamir,fursha,overlap,kikukawanoguch,borici}.
The 5D domain-wall operator is defined as

\beq
{\cal D} = \half \left\{ \gamma_5 \left( \partial^*_s + \partial_s \right) 
- a_s \partial^*_s \partial_s \right\} + {\cal M} \quad ,
\label{eq:dwf}
\eeq
where $s$ denotes a lattice site in the 5th direction
($1 \leq s \leq  N_s$),
$a_s$ is the corresponding lattice spacing, 
and $\partial^*_s$ and $\partial_s$ are 
the free forward and backward derivatives.

The operator ${\cal M}$ is obtained from 
the standard 4D Wilson--Dirac operator by
\beq
{\cal M } = D_W  - m_0
\eeq
with 
\beq
D_W = \half \left\{ \gamma_\mu \left( \nabla^*_\mu + \nabla_\mu \right) 
- a \nabla^*_\mu \nabla_\mu \right\} \quad .
\label{eq:wilson} 
\eeq
Here $\nabla^*_\mu$ and $\nabla_\mu$ are the gauge covariant 
forward and backward derivatives
and $a$ is the lattice spacing in the four physical dimensions 
$\mu=1,\dots,4$.
The domain-wall parameter $m_0$ obeys
\beq
0< a_s m_0 < 2 \quad , \quad 0 < a m_0 < 2 \quad.
\eeq
Note that the lattice spacings $a_s$ and $a$ can be different in general.
The boundary conditions in the DWF formulation in the 
5th direction is
\beq 
P_+ \psi(0,x) = P_- \psi(N_s + 1, x) = 0 \quad,
\eeq
where $P_\pm \equiv \half (1 \pm \gamma_5)$.
In these settings,
the chiral modes with opposite chiralities 
are localized on 4D boundary planes
at $s=1$ and $s=N_s$.

The 4D quark fields are constructed
from the left and right boundary (chiral) modes, as follows:
\beq
q(x) = P_- \psi(1,x) + P_+ \psi(N_s ,x)\quad ,  \quad
\bar{q}(x) = \bar{\psi}(1,x) P_+ + \bar{\psi}(N_s ,x) P_-  \quad .
\eeq

A bare quark mass term is introduced by adding to eq.~(\ref{eq:dwf}) the term
\beq
m_f \left\{ \bar{\psi}(1,x)P_+\psi(N_s ,x) 
+ \bar{\psi}(N_s ,x) P_-\psi(1,x) \right\} = m_f \bar{q}(x) q(x) \quad .
\eeq

The propagator of the quark fields is related to an effective 4D operator
$D_{N_s}$  \cite{overlap,kikukawanoguch,borici}
\beq 
\langle q(x) \bar{q}(x) \rangle = \frac{2- a D_{N_s}}{a D_{N_s,m_f} } 
\quad,
\eeq
with
\beq
D_{N_s,m_f} = \left(1- a m_f \right) D_{N_s} + 2 m_f \quad.
\eeq
In terms of the operators $K_\pm$,
\beq
K_\pm \equiv \half \pm \half \gamma_5 
\frac{a_s {\cal M}}{2 + a_s {\cal M}} \quad,
\eeq
$D_{N_s}$ is given by 
\beq 
a D_{N_s} = 1 + \gamma_5 \frac{K_+^{N_s} - K_-^{N_s}}{K_+^{N_s} + K_-^{N_s}} 
\quad.
\eeq
From this equation, it is easy to show that 
\beq
%a D \equiv \lim_{N_s \rightarrow \infty} a D_{N_s} 
%= 1 + \gamma_5 \epsilon \left( K_+ - K_- \right) \quad,
a D \equiv \lim_{N_s \rightarrow \infty} a D_{N_s} 
= 1 + \gamma_5 \, {\rm sign} \left( K_+ - K_- \right) \quad,
\eeq
which is written as
\beq
a D = 1 - \frac{A}{\sqrt{A^\dagger A}} \quad, 
\label{eq:gw} 
\eeq
\beq
A = - \frac{a_s {\cal M}}{2 + a_s {\cal M}} \quad.
\label{eq:matA}
\eeq
The operator $D$ in eq.~(\ref{eq:gw}) satisfies the Ginsparg--Wilson relation.
The only difference to Neuberger's operator \cite{neuberger} is the definition 
of $A$.
Neuberger's operator is obtained from eqs. (\ref{eq:gw}) and (\ref{eq:matA})
by taking the limit $a_s \rightarrow 0$.

In the limit $N_s \rightarrow \infty$, 
the 5D formulation of DWF is completely equivalent to 
a 4D lattice formulation of Ginsparg--Wilson fermions 
satisfying an exact chiral symmetry.
However, in a realistic simulation $N_s$ is kept finite, of course.
In this situation, the chiral symmetry is explicitly broken 
by the residual terms $\delta D \equiv D_{N_s} - D$.
A measurement of the effects of this chiral symmetry breaking 
is the so-called residual mass $m_\mathrm{res}$,
derived from the axial Ward--Takahashi identity.
The chiral transformation of DWF is defined as
\beq
\delta \psi(s,x) = i Q(s) \epsilon \psi(s,x) \,,\, \quad
\delta \bar{\psi}(s,x) =  - i \bar{\psi}(s,x) Q(s) \epsilon \,,
\eeq
where $Q(s) = {\rm sign}(N_s - 2 s + 1)$ 
and $\epsilon$ is an infinitesimal transformation parameter.
Under this transformation,
the quark fields are transformed as the usual chiral transformation:
$\delta q(x) = i \gamma_5 \epsilon q(x) \,,\,
\delta \bar{q}(x) = i \bar{q}(x)\gamma_5 \epsilon$ .
Therefore
the axial Ward--Takahashi identity is 
\beq
\sum_{\mu} \langle \nabla_\mu A_\mu(x) P(0) \rangle = 
2 m_f \langle P(x) P(0) \rangle + 2 \langle J_{5q}(x) P(0) \rangle 
\quad ,
\eeq
where $A_\mu(x)$ is the axial-vector current
and $P(x)$ is the pseudo-scalar density;
$A_\mu(x)$ and $P(x)$ are given as
\beq
A_\mu(x) =  \sum_s Q(s) J_\mu(s,x) \quad , \quad 
P(x) =  \bar{q}(x) \gamma_5 q(x) \quad , 
\eeq
where
\beq
J_\mu(x) = \frac{1}{2}\left[\bar{\psi}(s, x)(1-\gamma_\mu)U_\mu(x)
               \psi(s,x+\mu) - \bar{\psi}(s, x+\mu)(1+\gamma_\mu)
                U^\dagger_\mu(x) \psi(s,x)\right]  \quad .
\eeq

The additional term 
$J_{5q}$ represents the explicit breaking of chiral symmetry,
\beq
J_{5q} = - \bar{\psi}\left(\frac{N_s}{2}, x \right) P_- 
\psi \left(\frac{N_s}{2}+1,x \right) + 
\bar{\psi}\left(\frac{N_s}{2}+1,x\right) P_+ 
\psi\left(\frac{N_s}{2},x \right) \quad.
\eeq
For a smooth gauge field background,
the term $\langle J_{5q} P \rangle$  
vanishes exponentially fast \cite{fursha,kikukawa,locality}
as $N_s$ is increased.
In realistic simulations, however, the gauge fields can be rough and it
may happen that the rate of convergence in  $N_s$ is rather poor.
The breaking of chiral symmetry 
can be quantified 
by the values of $m_\mathrm{res}$.
Let us define the ratio $R(t)$: 
\beq 
R(t) = 
\frac{\sum_{\bf x} \langle J_{5q}({\bf x}, t) P({\bf 0}, 0) \rangle }{\sum_{\bf x} \langle P({\bf x}, t) P({\bf 0}, 0) \rangle} \quad.
\label{roft}
\eeq
The usual definition of $m_\mathrm{res}$ is 
the average of the quantity $R(t)$ 
at large time separations. 
A necessary but, maybe, not sufficient condition 
to recover fully on-shell chiral symmetry 
at a non-vanishing value of the lattice spacing
is that this quantity be negligible.  

%%%%%%%%%%%%%%%%%%%%%%%%%%%%%%% SECTION %%%%%%%%%%%%%%%%%%%%%%%%%%%%%%%%%%%%%
%\clearpage
\section{Eigenvalues of $A^\dagger A$}
\label{sec:eva2}

For gauge configurations with a restricted value of the plaquette
(so-called admissible configurations) \cite{kikukawa,locality},
the operator $A^\dagger A$ has been shown to have a spectral gap,
$0 < u \leq A^\dagger A \leq v$, 
ensuring the exponential suppression of the residual mass in $N_s$.
However, in realistic simulations as performed today,
the plaquette bound is not satisfied 
and it is important to study the distribution of 
the eigenvalues of $A^\dagger A$ in numerical simulations.

The eigenvalues of $A^\dagger A$ can be obtained 
through the generalized 4D eigenvalue equation \cite{numrec}
\beq
a_s^2 {\cal M}^\dagger{\cal M} \psi 
= \lambda \,\,(2 + a_s {\cal M})^\dagger (2 + a_s {\cal M}) \psi \quad.
\eeq
The low-lying (maximal) eigenvalues can be computed by 
minimizing (maximizing)
the generalized Ritz functional\footnote{The interested reader may obtain 
more details on request.}
\beq
\frac{\langle \psi | a_s^2 {\cal M}^\dagger{\cal M} | \psi \rangle}{
\langle \psi | (2 + a_s {\cal M})^\dagger (2 + a_s {\cal M})| \psi \rangle}
\label{eq:ritz}
\eeq
using a straightforward generalization of the algorithm described 
in ref. \cite{ritz}. 
Notice that in this method no inversion of the matrix 
$(2 + a_s {\cal M})^\dagger (2 + a_s {\cal M})$ is needed.
Higher eigenvalues can be calculated 
by modifying the operator ${\cal M}^\dagger{\cal M}$
in the numerator in eq.~(\ref{eq:ritz}), so that     
the already computed eigenvalues are shifted to larger values.
This can be achieved \cite{project} by substituting 
\beq
{\cal M}^\dagger{\cal M} \rightarrow {\cal M}^\dagger{\cal M}
+ \sum_i \left( \frac{1-\lambda_i}{\lambda_i} \right)
{\cal M}^\dagger{\cal M} |\psi_i \rangle \langle \psi_i |
(2 + a_s {\cal M})^\dagger (2 + a_s {\cal M}) \,,
\eeq
$\lambda_i$, $\psi_i$ being the already computed 
(lower) eigenvalues and eigenvectors.

\begin{figure}[h] 
\centering
\centerline{
\resizebox{10.5cm}{!}{\rotatebox{-90}{\includegraphics{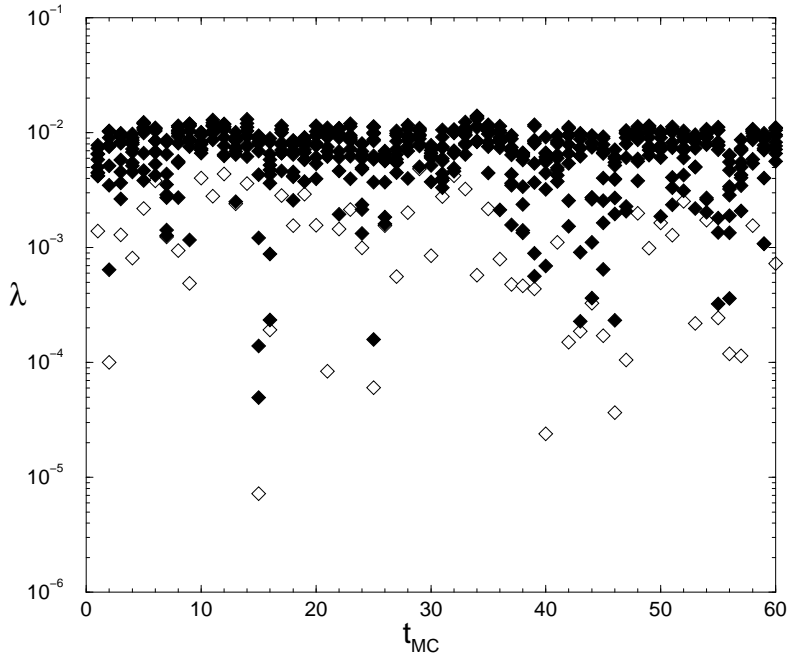}}}
}
\caption{
The 11 lowest eigenvalues
of the operator $A^\dagger A$ 
as a function of Monte Carlo time $t_\mathrm{MC}$ 
at $\beta=6.0$ and $m_0=1.8$ on a $12^3 \times 24$ lattice.
The open diamonds denote the lowest eigenvalue.
}
\label{fig:ev2}
\end{figure}

Figure \ref{fig:ev2} shows the eleven lowest eigenvalues 
of the operator $A^\dagger A$ 
as a function of the Monte Carlo time ($t_\mathrm{MC}$).
Here and throughout the paper 
we use the quenched approximation and set $a_s=1$. 
The data in Fig.~\ref{fig:ev2} are obtained 
with the Wilson gauge 
action at
$\beta=6.0$ 
on a $12^3 \times 24$ lattice,
setting $m_0=1.8$.
As expected, very small eigenvalues appear frequently.

%\begin{figure}[h!tb] 
%\begin{figure}[ht] 

\begin{figure}[ht] 
\centering
\centerline{
\resizebox{10.5cm}{!}{\rotatebox{-90}{\includegraphics{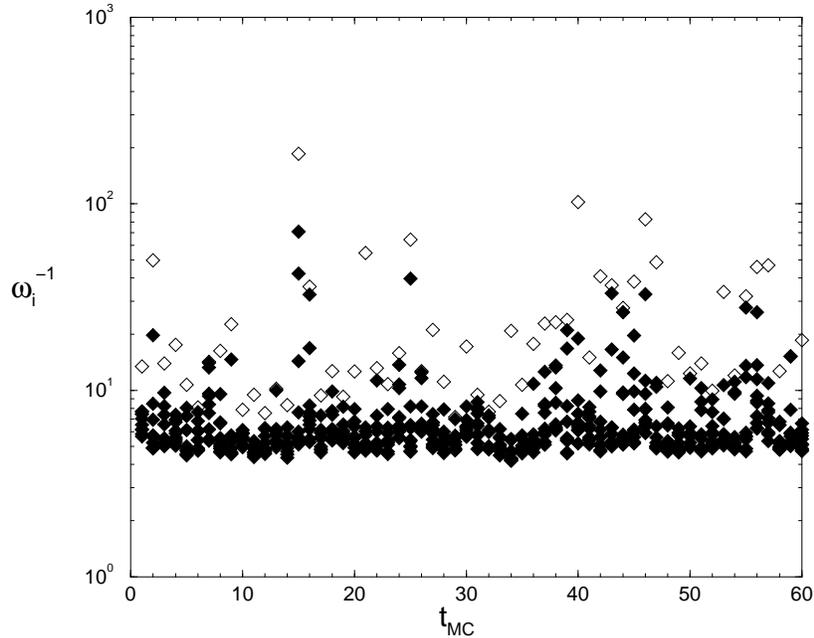}}}
}
\caption{
The inverse convergence rate $\omega_i^{-1}$ 
computed from the eigenvalues plotted in Fig~\ref{fig:ev2}.
}
\label{fig:convr}
\end{figure}

The minimum rate of convergence in $N_s$ of the operator 
$D_{N_s}$ is given by
\beq 
\omega = \min_i [\omega_i] \quad, \quad
\omega_i \equiv \ln \frac{1 + \sqrt{\lambda_i}}{| 1 - \sqrt{\lambda_i}|} 
\quad ,
\eeq
where $\lambda_i$ are the eigenvalues of $A^\dagger A$ \cite{project}.
Figure \ref{fig:convr} shows the inverse convergence rate
computed from the eigenvalues in Fig.~\ref{fig:ev2}.
Clearly, the low-lying eigenvalues of $A^\dagger A$ lead to a 
slow convergence, 
causing the simulation to become expensive.

%\begin{figure}[h!tb] 

\begin{figure}[ht] 
\centering
\centerline{\resizebox{10.5cm}{!}
{\includegraphics{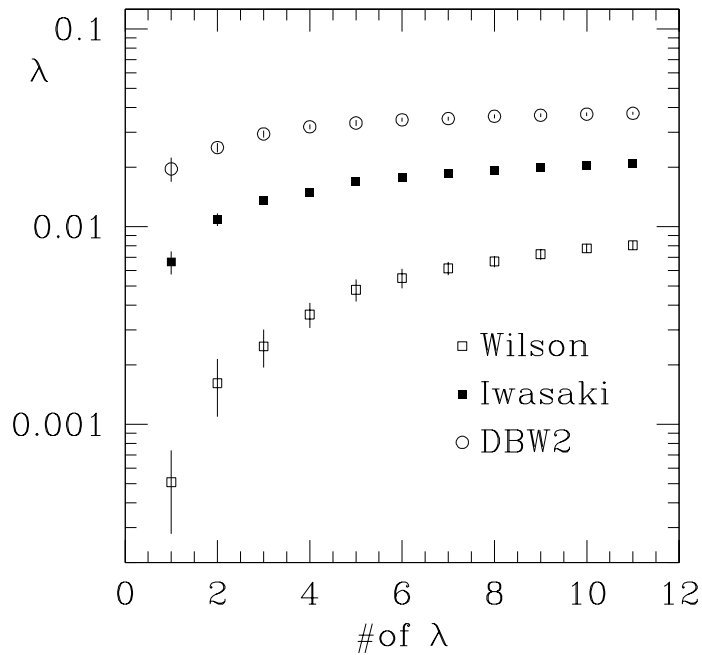}}}
\caption{Averaged eigenvalues for the Wilson, Iwasaki and DBW2 gauge actions
as a function of the eigenvalue number. 
The lattice spacing 
$a=0.093$ fm used is 
the same for all gauge actions.                              
}
\label{fig:gauge}
\end{figure}

We also explored the eigenvalues for other gauge actions such as
the Iwasaki \cite{iwasakiaction} and DBW2 \cite{dbw2action} ones. 
An example for these eigenvalues is plotted in Fig.~\ref{fig:gauge}. 
%There 
In that figure
we average over 20 gauge configurations. 
The parameters of the gauge actions were 
chosen such that in each case the value of the lattice spacing 
is $a=0.093$ fm, 
leading to setting $\beta=6.0$ for the Wilson action, 
$\beta=2.6$ for the Iwasaki one 
and $\beta=1.04$ for the DBW2 one. 
Since also the lattice size was fixed to 
be $12^3\times 24$ we have for the different gauge actions the same 
physical situation. 
For the Wilson action we observe small values 
for the lowest-lying modes. 
This is improved substantially 
by employing the Iwasaki action and even more when using the DBW2 action. 
Note that the 11th low-lying eigenvalue of the Wilson action 
corresponds to the lowest eigenvalue of the Iwasaki action.  
We checked for the Wilson and the Iwasaki action that this picture
does not change when we decrease the value of the lattice spacing 
down to $a=0.05$ fm. 
This confirms that the convergence in $N_s$ is faster 
when the gauge action is improved \cite{cppacs,rbc,dbw2}.  
As we will see below
a conclusion that improved gauge actions by themselves 
would completely cure the problem of a slow convergence rate 
is premature, however. 

%\begin{figure}[h!tb] 

%%%%%%%%%%%%%%%%%%%%%%%%%%%%%%% SECTION %%%%%%%%%%%%%%%%%%%%%%%%%%%%%%%%%%%%%
\section{Improvement of domain-wall fermion}
\label{sec:impdwf}

The decay rate of the residual mass in $N_s$ is controlled
by the small eigenvalues of $A^\dagger A$.
For the Wilson gauge action very small eigenvalues occur, 
leading to a slow convergence. 
Although the situation is improved 
for the Iwasaki gauge action,
as we saw above, it was observed that even
in this case for large values of $N_s$ the convergence turned to become
very slow \cite{cppacs,eigenv}. 
It thus seems to be necessary to test methods as proposed in 
\cite{project} that 
modify the fermionic part of the DWF action  
by projecting out the small eigenvalues of $A^\dagger A$.
These methods 
can be used alternatively --or even in addition-- 
to employing improved gauge actions.                 
The key observation in \cite{project} is that 
the relations in eqs.~(\ref{eq:gw}) and (\ref{eq:matA})
hold true for any choice of ${\cal M}$ as long as
\beq
{\cal M}^\dagger = \gamma_5 {\cal M} \gamma_5 \quad, \quad
\det(2 + a_s {\cal M}) \neq 0 \quad.
\eeq
This fact may be used to construct an improved ${\cal M}$
for which the very low eigenvalues of $A^\dagger A$ disappear.

Let us, for completeness, 
repeat the construction of the improved 
operator here again following \cite{project}. 
The basic idea is to find the new operator ${\widehat {\cal M}}$ 
satisfying the following relation;
\beq
\widehat A = - \frac{a_s \widehat {\cal M}}{ 2+ a_s {\widehat {\cal M}}}
= A + \sum_{k=1}^r (\widehat \alpha_k - \alpha_k ) 
\gamma_5 v_k \otimes v_k^\dagger \quad ,
\label{ahat}
\eeq
where $v_k$ is the eigenvector of the following equation
\beq
\gamma_5 A v_k = \alpha_k v_k \quad, \quad k=1,\dots,r \quad,\quad
(v_k, v_l) = \delta_{k l} \quad.
\eeq
Therefore an improved DWF operator, $D_\mathrm{dwf}^\mathrm{imp}$, 
can be obtained from  eq.~(\ref{eq:dwf}) after substituting
${\cal M}$ with $\widehat {\cal M}$ defined as
\beq 
a_s \widehat {\cal M} = a_s {\cal M} - \sum_{k,l=1}^r
X_{kl} w_k \otimes w_l^\dagger \gamma_5 \quad,
\eeq 
where 
\beq
w_k = (2 + a_s {\cal M}) \gamma_5 v_k 
\eeq
and 
\beq
(X^{-1})_{kl} = 2 \delta_{kl} (\widehat \alpha_k - \alpha_k)^{-1}
+ (v_k,w_l) \quad.
\eeq
It is easy to see that $\gamma_5 \widehat A$ has the same eigenvectors
as $\gamma_5 A$;
however, all eigenvalues $\alpha_k$, $k=1,\dots,r$, are replaced by 
${\widehat\alpha}_k$. 
The limit $N_s\rightarrow \infty$ of $D_{N_s}$ is 
of course unchanged by this modification, provided
\beq
{\rm sign}({\widehat \alpha}_k) = {\rm sign}(\alpha_k) \,. 
\eeq

The choice of $|{\widehat \alpha}_k|$ is not unique. We will choose here
\begin{equation}
{\widehat \alpha}_k = 2\; {\rm sign}(\alpha_k) |\alpha_l| \,,\quad 
1 \le k\le r\equiv k_\mathrm{max} \, ,
\label{alphachoice}
\end{equation}
where $k_\mathrm{max}$ is the number of eigenvalues projected out
and $l$ can be chosen freely. A natural
choice is $l=k_\mathrm{max}$ such that all low-lying eigenvalues are moved
to be twice higher than the largest eigenvalue projected out. 
We also tried, however, different values of $l$ 
and found that the improvement is not very sensitive to 
the precise choice of ${\widehat \alpha}_k$, 
provided it is larger than $\alpha_{k_\mathrm{max}}$. 

Our statistics is typically $60$ configurations for the Wilson 
gauge action and $20$ configurations for the Iwasaki action.
We did not explore the DBW2 action extensively.
The parameters of the gauge actions were chosen as before such that 
$a^{-1}=2$ GeV, 
which means a choice of $\beta=6.0$ for 
the Wilson gauge action 
and $\beta=2.6$ for the Iwasaki one.
The lattice sizes were $12^3 \times 24 \times N_s$ and 
$16^3 \times 24 \times N_s$ for the two actions, respectively. 
The domain-wall mass was $m_0=1.8$ and we worked at a quark mass of 
$m_f=0.02$.

%\begin{figure}[h!tb] 
%\begin{figure}[ht] 
\begin{figure}[h] 
\centering
\centerline{\resizebox{10.5cm}{!}
{\rotatebox{-90}{\includegraphics{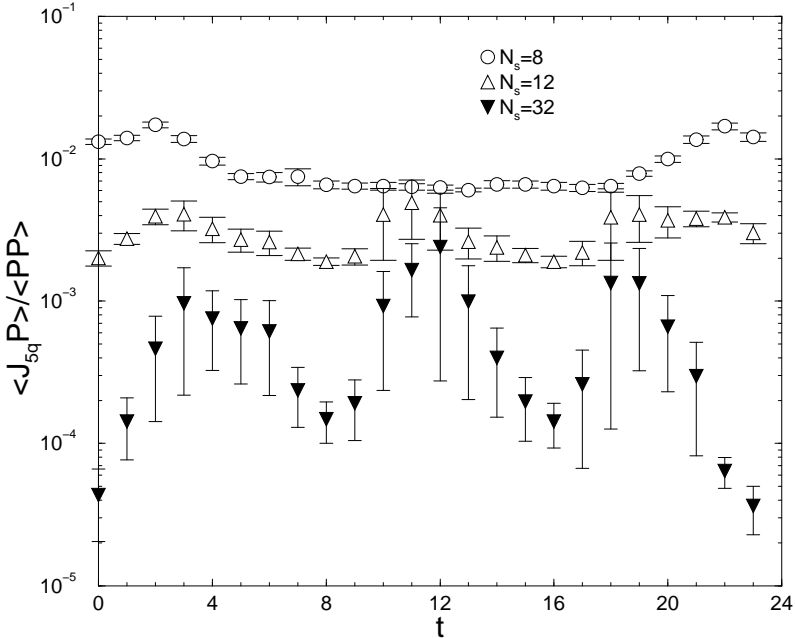}}}}
\caption{The ratio
$R(t)=\frac{\langle J_{5q}P \rangle}{ \langle PP \rangle}$
as a function of Euclidean time. The Wilson gauge action 
is chosen with simulation parameters as given in the text.
No projection of eigenvalues is performed. 
}
\label{fig:proj0}
\end{figure}

We have measured the residual mass from $R(t)$ in 
eq.~(\ref{roft}) as the average of $R(t)$ for $t$ typically
in the interval $4 \le t \le 20 $ 
for a time extent of the lattice of $T=24$.
$R(t)$ is shown in 
Fig.~\ref{fig:proj0} for the case 
when no projection is performed. 
For each value of $N_s$ we have the same statistics.
Although, with increasing $N_s$, 
the residual mass $m_\mathrm{res}$ decreases, 
it does so rather slowly; 
furthermore, as $N_s$ increases, 
large fluctuations in $R(t)$ occur, 
rendering the determination of the residual mass difficult. 
These large fluctuations also suggest 
that the residual chirality-breaking effects 
in other quantities might be very hard to estimate,
taking only $m_\mathrm{res}$ as a measure of these effects. 

\begin{figure}[h]
\centering
\centerline{\resizebox{10.5cm}{!}
{\rotatebox{-90}{\includegraphics{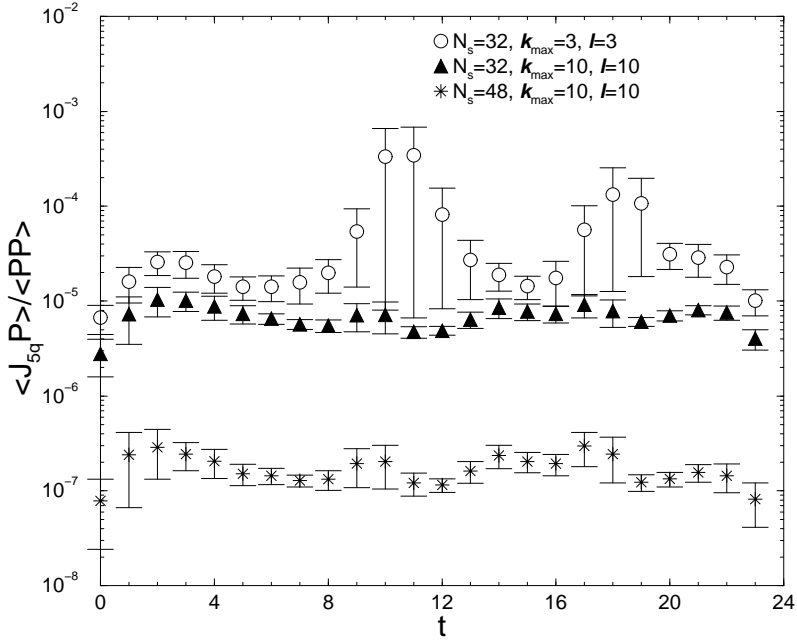}}}}
\caption{Same as Fig.~\ref{fig:proj0} but now with the projection of
eigenvalues employed.
}
\label{fig:projns32}
\end{figure}

In Fig.~\ref{fig:projns32} we show $R(t)$ 
when we project out a number of eigenvalues. 
As expected, the projection of the low eigenvalues 
decreases the residual mass significantly 
with respect to Fig.~\ref{fig:proj0}.

%\begin{figure}[h!tb]
%\begin{figure}[ht]

\begin{figure}[h] 
\centering
\centerline{\resizebox{10.5cm}{!}
{\rotatebox{-90}{\includegraphics{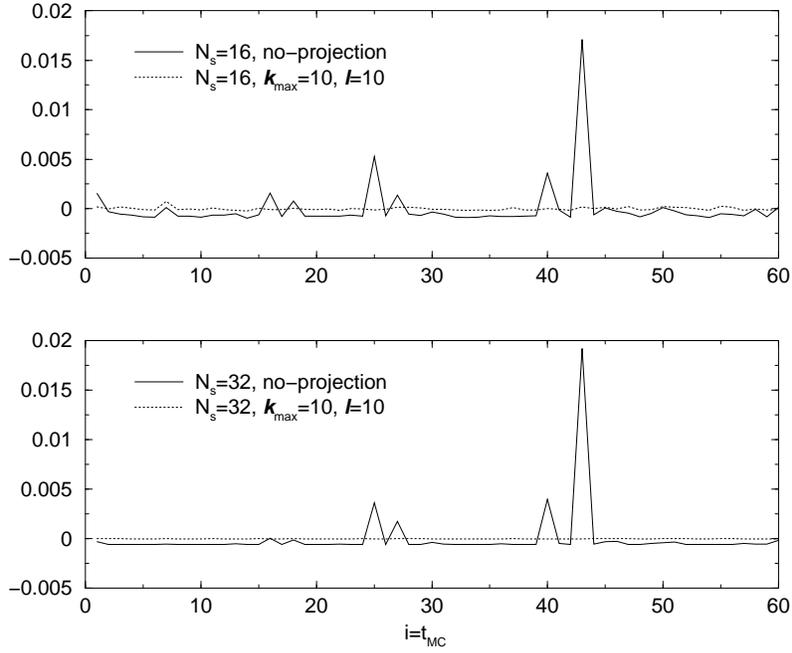}}}}
\caption{
The quantity $\widehat{R}(t) - \langle \widehat{R}(t) \rangle$ at $t=12$,
see eq.~(\ref{rhat}), 
as a function of Monte Carlo time for the Wilson gauge action, 
with and without projection. 
}
\label{fig:mc_histo_wilson}
\end{figure}

The more eigenvalues are projected the smaller the residual mass is. 
Another important feature is 
that the fluctuations in $R(t)$ become much smaller 
when a sufficiently large number of eigenvalues is projected out; 
in this case 10 seems to be a good choice.
This is very clearly seen in Fig.~\ref{fig:mc_histo_wilson}, 
where we show the value of the ratio $\widehat{R}(t)$, 
\beq
\widehat{R}(t) = \frac{\sum_{\bf x} J_{5q}({\bf x}, t) P({\bf 0}, 0)}
{\sum_{\bf x} P({\bf x}, t) P({\bf 0}, 0)} \quad,
\label{rhat}
\eeq
computed on single configurations
at $t=12$ as a function of Monte Carlo time. 
The spikes are 
substantially damped with the projection.   
%\begin{figure}[h!tb] 
%\begin{figure}[ht] 
Finally, when the projection is implemented, the decrease of the residual 
mass with $N_s$ is much faster. 

In summary, it is clear that the projection 
method has a drastic effect on the value and dispersion of the 
residual mass. However a sufficient ($O(10)$ in our setup) number of 
eigenvalues have to be projected out.

%%%%%%%%%%%%%%%%%%%%%%%%%%%%% SECTION %%%%%%%%%%%%%%%%%%%%%%%%%%%%%%%%%%%
\section{DWF with improved gauge actions}
\label{sec:impgauge}

\begin{figure}[h] 
\centering
\centerline{\resizebox{7cm}{!}{\rotatebox{-90}
           {\includegraphics{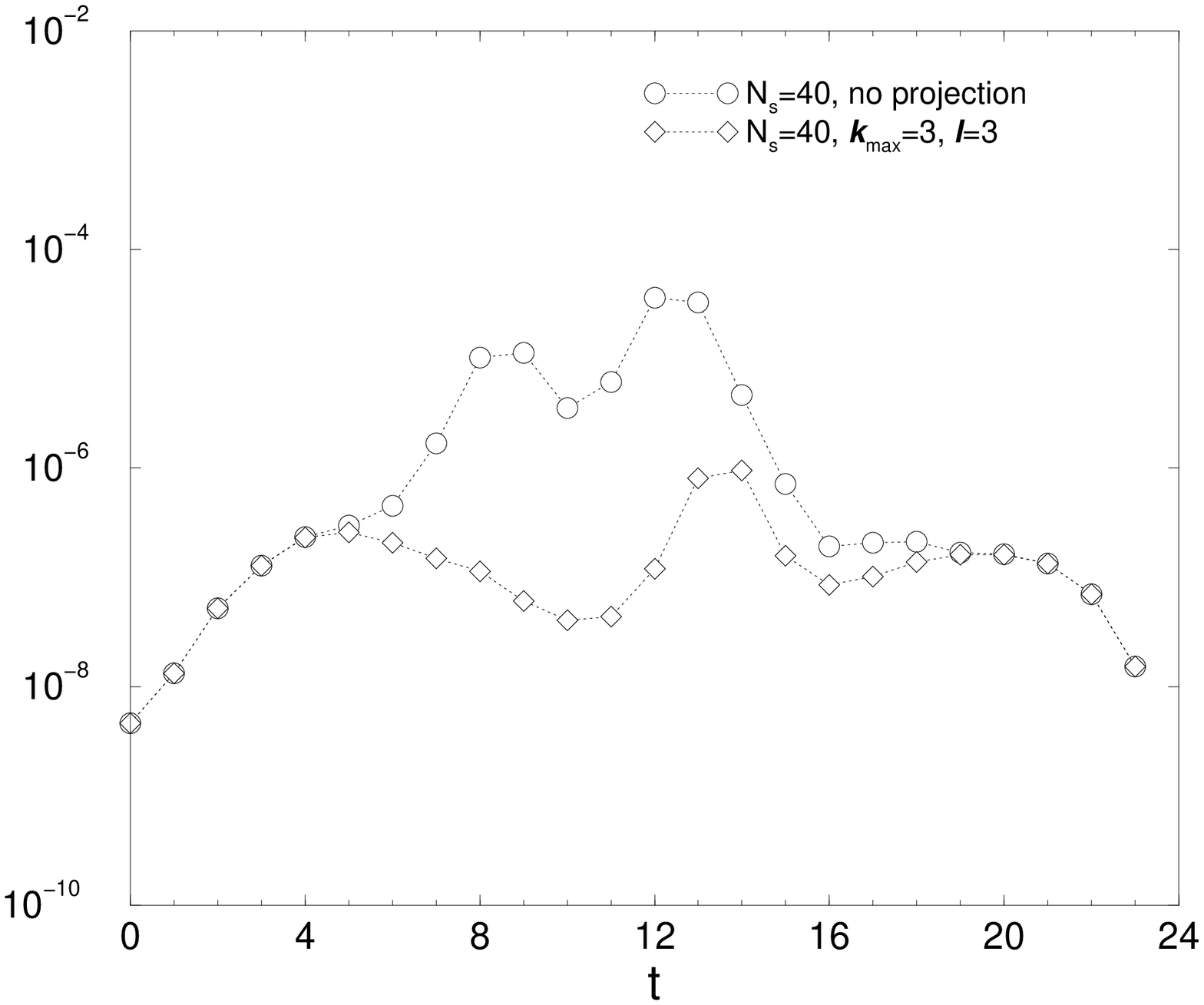}}}
            \resizebox{7cm}{!}{\rotatebox{-90}
           {\includegraphics{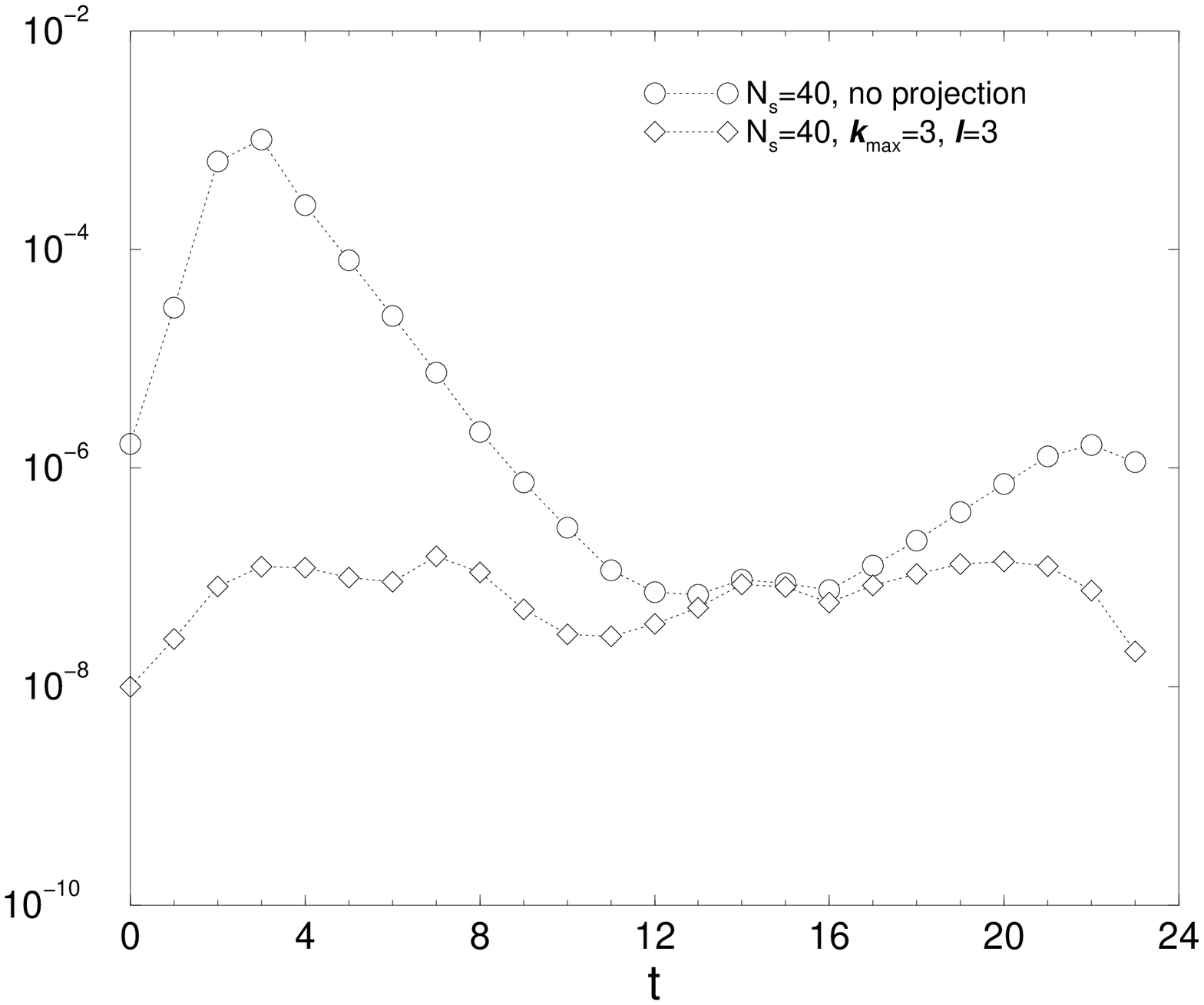}}}}
\caption{
$\widehat{R}(t)$ as a function of time in the Iwasaki action.
The circles show the results for the no-projection case 
and the diamonds for the case when 3 eigenvalues are projected out.
}
\label{fig:rg}
\end{figure}

\begin{figure}[h] 
\centering
\centerline{\resizebox{10.5cm}{!}
{\rotatebox{-90}{\includegraphics{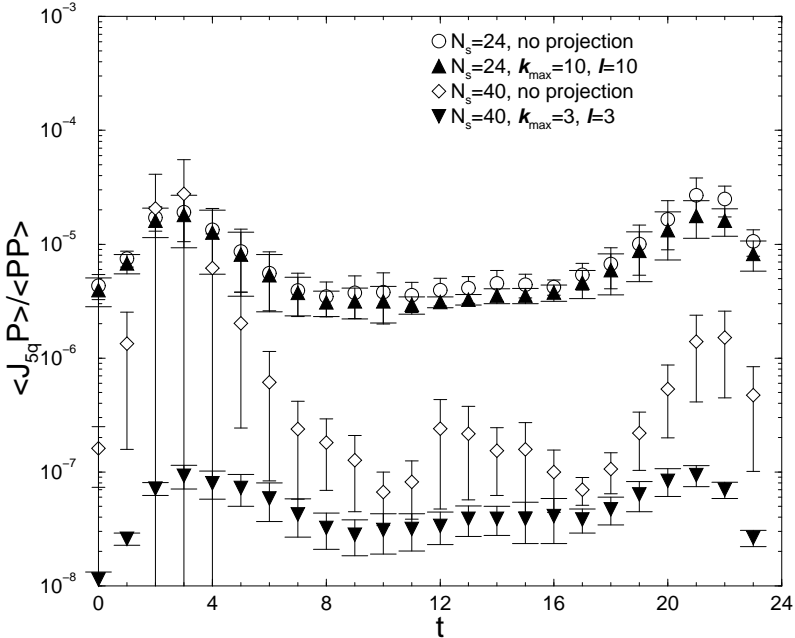}}}}
\caption{
$\frac{\langle J_{5q}P \rangle}{ \langle PP \rangle}$
without and with projection for the 
Iwasaki gauge action.
}
\label{fig:rgn40}
\end{figure}

In our simulations with the Iwasaki gauge action we found, 
even in our small sample of only 20 configurations,
very low-lying eigenvalues of $A^\dagger A$. 
In order to see the effect of these modes we plot, 
in Fig.~\ref{fig:rg},
the ratio $\widehat{R}(t)$ of eq.~(\ref{rhat}),                         
for two of these configurations 
(note that no averaging is involved here). 
The figures indicate that 
we will find, 
also for the Iwasaki gauge action,
the same problem as for the Wilson gauge action. 
When no projection is performed, 
the correlation function shows a spiky behaviour,
which may lead to large fluctuations 
in $\widehat{R}(t)$ and hence to a very difficult determination 
of the residual mass.
This is also confirmed  
in the ratio of averaged values, $R(t)$, 
as shown in Fig.~\ref{fig:rgn40}. 
The pattern resembles the case of the Wilson gauge action. 
For small values
of $N_s$ the effect of the projection is not noticeable. 
For larger values of $N_s$, we see that $R(t)$ is lowered when 
the eigenvalues are projected out and that the fluctuations of this quantity 
are strongly damped. 
%\begin{figure}[h!tb] 
%\begin{figure}[ht] 

%\begin{figure}[h!tb] 
%\begin{figure}[ht] 
We also made an attempt to see how the projection method
affects $R(t)$ for the DBW2 action. 
For the simulations 
we chose $\beta=1.04$, 
which corresponds again to $a^{-1}=2$ GeV.
Thus we study the same physical situation 
%as before 
with the Wilson and Iwasaki gauge actions. 
The lattice size was chosen 
to be $16^3\times 32\times N_s$ and $m_0=1.7$. 
The fermion mass was taken to be $m_f=0.02$.

\begin{figure}[h] 
\centering
\centerline{\resizebox{10.5cm}{!}
{\rotatebox{-90}{\includegraphics{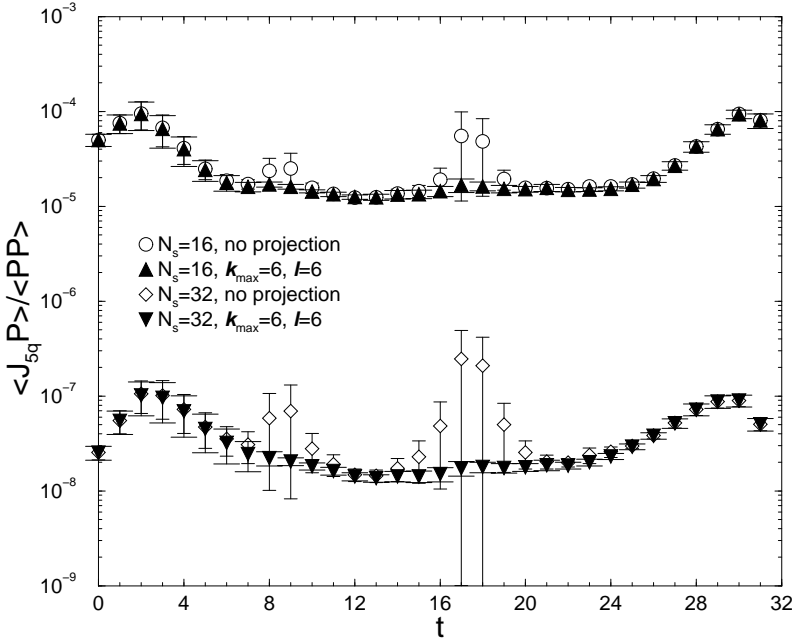}}}}
\caption{
$\frac{\langle J_{5q}P \rangle}{ \langle PP \rangle}$
without and with projection at $N_s=16$ and 32,
for the DBW2 gauge action.
}
\label{fig:dbw2res}
\end{figure}

We observe in Fig.~\ref{fig:dbw2res} that the residual mass is not 
changed very much by the projection. We attribute this
to the fact that in our small statistical sample no very low-lying
eigenvalues of $A^\dagger A$ could be detected. 
We see, however, from 
%Fig.~\ref{fig:dbw2res} 
the same figure 
that the statistical error is substantially 
reduced for certain values of $t$ when the projection of eigenvalues 
is employed.  

%\begin{figure}[h!tb] 
%\begin{figure}[ht] 

The fact that $R(t)$ shows large fluctuations, 
even though there are no very small low-lying eigenvalues,
points toward the suspicion that also 
the eigenvectors may play an important role. 
In particular the localization properties of these eigenmodes 
may lead to large fluctuations 
as discussed in \cite{aokitani}.
Although this point deserves 
further investigation, we did not perform such a study here. 
To conclude,
from a negligible average value of the residual mass, 
that chiral symmetry is restored is certainly questionable 
when the dispersion of the residual mass is large and not gaussian. 
A much safer situation would be to ensure that the residual mass is bounded
from above for {\it all} configurations. The projection method ensures 
that this is the case. 

\begin{figure}[ht] 
\centering
\centerline{\resizebox{10.5cm}{!}
{\rotatebox{-90}{\includegraphics{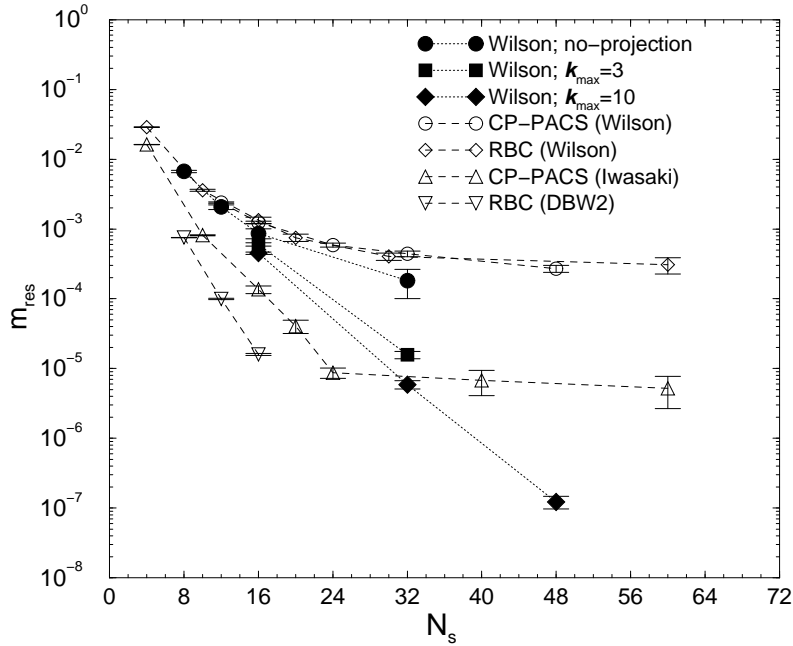}}}}
\caption{A compilation of the residual mass as a function of $N_s$
for various gauge actions and various choices of projecting 
eigenvalues. The filled symbols correspond to our own results.
The data of the DBW2 action are taken from \cite{rbc} 
and the ones for the Iwasaki action from \cite{cppacs,eigenv}.
The lines are just to guide the eye.
}
\label{fig:resmass}
\end{figure}

To summarize, in Fig.~\ref{fig:resmass} 
we show 
the comparison of the behaviour of the residual mass 
as a function of $N_s$ for different gauge actions and 
for different numbers of projected eigenvalues. 
For a fixed gauge action, 
we find 
that at small $N_s$
there is almost no effect from the projection method. 

This can be explained by a simple qualitative argument
with the formula suggested in \cite{shamirpert,eigenv,aokitani};
\beq
m_\mathrm{res} \sim \sum_k e^{-\alpha_k N_s}
\sim \int d \alpha \rho(\alpha) e^{-\alpha N_s} \quad ,
\label{prediction}
\eeq
where  
$\rho(\alpha)$ is the eigenvalue density in the continuum.
This (qualitative) formula describes the behaviour of $m_\mathrm{res}$ 
as a function of $N_s$. 
The formula contains two factors, the eigenvalue density and the 
exponential supression factor $e^{- \alpha N_s}$.  
For small values of $N_s$, 
not only do the low-lying modes
contribute to the sum in eq.~(\ref{prediction}),
but also the bulk modes since they are not supressed sufficiently.
When projecting out a few number of low-lying eigenmodes, 
the eigenvalue density and the exponential factor
remain almost unchanged and hence also  
the residual mass is not affected very much for small 
values of $N_s$. 
In such a case, it would be necessary to project out
a large number of eigenmodes 
to make $m_\mathrm{res}$ decrease.
When $N_s$ is chosen to be large, 
on the other hand,
the bulk mode contributions
to the sum in eq.~(\ref{prediction}) will die out 
and only the small eigenvalue contributions will survive. 
As a consequence, the factor $e^{-\alpha N_s}$ becomes much smaller
after projecting out even only a few low-lying (isolated) eigenmodes.
This should hence lead to a large improvement, i.e. a substantial 
decrease of the residual
mass when the projection method is active. 
As Fig.~\ref{fig:resmass} clearly shows, this is indeed the case.
For the Wilson gauge action at $N_s=48$, 
the value of the residual
mass is decreased by several orders of magnitude when 10 eigenvalues
are projected out. 
We made a rough check for the Iwasaki gauge action 
that also in this case the residual mass
decreases substantially, choosing $N_s=40$.  
Thus the very slow decrease of the residual mass as a function of $N_s$ 
in the original DWF formulation with no projection 
is cured by projecting out a few $O(10)$ eigenvalues. 

%\begin{figure}[h!tb] 

%%%%%%%%%%%%%%%%%%%%%%%%%%%%%%% SECTION %%%%%%%%%%%%%%%%%%%%%%%%%%%%%%%%%%%%%
\section{Conclusion}
\label{sec:conclusion}

We have studied the effect of modifying the fermion action
of DWF by projecting out a few low-lying eigenvalues of the 
underlying transfer matrix \cite{project}. 
By measuring the correlation function leading to a 
determination of the residual mass and the residual mass itself
as a function of $N_s$, 
we find a significant improvement 
in the restoration of chiral symmetry for quenched DWF at large $N_s$.

The reason is that 
in the large-$N_s$ limit
the low-lying eigenvalues of $A^\dagger A$ are responsible for 
the exponential convergence rate of DWF in $N_s$ to its chiral invariant 
limit. These eigenvalues then 
dominate the behaviour of the 
residual mass and whenever very small low-lying modes appear they 
lead to a very slow decrease of the residual mass 
as $N_s$ is increased.  
Projecting out a small number of these modes can therefore help
considerably to lower the values of the residual mass. 
We have confirmed
this picture in practical simulations, 
using the Wilson and the Iwasaki gauge actions. 
We observe that when a sufficient number, i.e. 
$O(10)$, eigenvalues are projected out, the residual mass vanishes 
rapidly with increasing $N_s$. 

Let us end our discussion with three remarks. \\
$(i)$ 
Projecting out a number of low-lying eigenvalues shows 
a strong effect not only on the value but also on the fluctuations 
of the correlation function $R(t)$ in eq.~(\ref{roft}) 
and hence of the residual mass. 
The damping of the fluctuations takes place even when 
no very small eigenvalues occur in the simulation,
as in the case of the DBW2 action.
It thus seems that also the eigenvectors and in particular
their localization properties play an important role. 
It is unclear to us, and we did not investigate this here, 
how far also other correlation functions are affected by this phenomenon. 
One possible explanation \cite{aokitani} relies
on the relation of the eigenvalues and eigenvectors
of $D_W$ and $D_{N_s, m_f}$.
The study of this correspondence clearly deserves
further efforts using non-perturbative methods.

\noindent $(ii)$ 
The method of projecting out eigenvalues as studied here 
can be used on top of other improvements 
such as using improved gauge actions or improved fermion actions. 
The projection method is not very costly and 
produces only a small numerical overhead. 
Thus we advocate to employ the projection method in any simulation 
done with DWF. 

\noindent $(iii)$ 
We expect that the projection of the low-lying eigenvalues
should play an even more
important role in the case of dynamical simulations with DWF 
as the behaviour of the 5D fermionic kernel 
will be affected by the problems discussed in $(i)$, too.
We envisage that such a dynamical computation 
with the projection of low-lying eigenvalues can be performed along
the lines of refs.~\cite{liu,boricidyn,ahasen,wolff} by estimating 
the full DWF operator stochastically. In this case
the projection can be done easily.

%%%%%%%%%%%%%%%%%%%%%%% ACKNOWLEDGMENTS %%%%%%%%%%%%%%%%%%%%%%%%%%%%%%%%%
\section*{Acknowledgment}
We are indebted to Pilar Hern\'andez for many valuable discussions and
suggestions. We gratefully acknowledge her contributions in an early
stage of the project.
We are most grateful to Silvia Necco 
for providing us with the update programme for the improved gauge actions.
We thank the John von Neumann institute for computing 
for providing the necessary computer time for this work.
K.-I.N. is supported by Japan Society for the Promotion of Science (JSPS)
Fellowship for Research Abroad.
This work is supported in part by the European Union
Improving Human Potential Programme
under contracts No.\ HPRN-CT-2000-00145 (Hadrons/Lattice QCD)
and HPRN-CT-2002-00311 (EURIDICE).

%-------------------------------------------------------------------

\end{document}